# Enhancing the sensitivity of mesoscopic light reflection statistics in weakly disordered media by interface reflections

**Daniel J. Park,[1,2] Prabhakar Pradhan,[3*] and Vadim Backman[1]**

[1]*Biomedical Engineering Department, Northwestern University, Evanston, IL 60208 USA*
[2]*Department of Chemistry, Northwestern University, Evanston, IL 60208 USA*
[3]*Department of Physics, BioNanoPhotonics Lab, University of Memphis, Memphis, TN 38152 USA*
[*]*ppradhan@memphis.edu*

**Abstract:** Reflection statistics have not been well studied for optical random media whose mean refractive indices do not match with the refractive indices of their surrounding media. Here, we theoretically study how this refractive index mismatch between a one dimensional (1D) optical sample and its surrounding medium affects the reflection statistics in the weak disorder limit, when the fluctuation part of the refractive index ($\Delta n$) is much smaller than the mismatch as well as the mean refractive index of the sample ($\Delta n << <n>$). In the theoretical derivation, we perform a detailed calculation that results in the analytical forms of mean and standard deviation (STD) of the reflectance in terms of disorder parameters ($\Delta n$ and $l_c$) in an index mismatched backscattering system. Particularly, the orders of disorder parameters in STD of the reflectance for index mismatched systems is shown to be lower ( $\sim(<\Delta n^2> l_c)^{1/2}$ ) than that of the matched systems ($\sim <\Delta n^2> l_c$). By comparing STDs of the reflection coefficient of index matched and mismatched systems, we show that reflectance at the sample boundaries in index mismatched systems can enhance the signal of the STD to the "disorder parameters" of the reflectance. In terms of biophotonics applications, this result can lead to potential techniques that effectively extract the sample disorder parameters by manipulating the index mismatch conditions. Potential applications of the technique for enhancement in sensitivity of cancer detection at single cell level are also discussed.

## 1. Introduction

The statistical properties of transport in 1D mesoscopic optical and electronic disordered media have been well studied [1-5]. The quantum mechanical and optical systems are analyzed within the same formalisms because the Schrödinger equation and Maxwell's equations are reduced to the same Helmholtz equation [6-9]. After the Landauer formalism showed that the reflection coefficient is related to the resistance or conductance of a sample, the outward scattering information such as reflection and transmission coefficients became important for the studies of localization and conductance fluctuations in electronic systems [6,7]. In optical research such as light scattering and localization, the disorder properties of the refractive index have been analyzed mainly under the assumption that the mean of a sample refractive index matches with the refractive index of its surrounding medium [6,8-10]. The results based on this assumption show that the mean and fluctuation (STD) of reflection coefficient (r), $\sigma(r)$, have the same analytical form in the weak disorder limit ($\Delta n << <n>$). However, reflection statistics have not been thoroughly studied for systems where the mean of the sample refractive index does not match with the refractive index of the surrounding medium.

The STD of reflectance exhibits significantly different behaviors in index mismatched systems in the weak disorder limit. The orders of disorder parameters ($\Delta n$ and $l_c$) in the mean of the reflectance are the same in the matched and mismatched systems ($<r> \sim <\Delta n^2>l_c$). But the STD of reflectance in mismatched systems deviates from that of matched systems, having different orders of disorder parameters ($\sigma(r) \sim <\Delta n^2>^{1/2}l_c^{1/2}$).

In this paper, we present a detailed theoretical derivation and a physical interpretation of reflection statistics in index mismatched systems. In section 2, by applying the index mismatch condition, and we re-express the Langevin equation with two separate equations: a deterministic equation and a stochastic equation. Then, a short-range correlated random noise is introduced to the refractive index of an optical sample in the weak disorder limit. Finally, the mean and STD of the reflectance are calculated by the perturbative expansions in terms of disorder parameters and boundary index mismatch terms. In section 3, based on the results in section 2, STDs of the reflectance in matched and mismatched systems are compared, and the physical origin of their difference is explained. A potential application of the technique in mismatched systems is also discussed in the context of efficiently extracting disorder parameters in more efficient way in biological cell systems from the back scattering signal.

## 2. Theoretical calculation

*2.1 Langevin equation with boundary index mismatch condition*

In this section, we construct the framework of theoretical derivation by reformulating the stochastic equation for the reflection coefficient with the refractive index mismatch condition. For this, let us first consider a 1D optical dielectric random medium that exists in $0 < x < L$ and its surrounding medium ($x > L$ or $x < 0$). In the stationary regime, a wave function $\psi(x)$ of an electromagnetic field is described by the Maxwell's equation:

$$\frac{d^2\psi}{dx^2} + k^2 n^2(x)\psi = 0, \qquad (1)$$

where $k$ is a wave vector in a vacuum and $n(x)$ is the medium refractive index in $0 < x < L$.

When a plane wave with a wave vector $k$ impinges on the random medium from the right as shown in Fig. 1, the wave function has a form: $\psi(x) = e^{-ik(x-L)} + R(L)e^{ik(x-L)}$ ($x > L$) and $\psi(x) = T(L)e^{-ikx}$ ($x < 0$), where $R(L)$ and $T(L)$ are reflection and transmission coefficients. Based on the invariant imbedding approach [6], $R(L)$ is described by a stochastic differential equation (Langevin equation) with an initial condition:

$$\frac{dR(L)}{dL} = 2ikR(L) + \frac{ik}{2}\eta(L)[1+R(L)]^2, \qquad (2a)$$
$$R(L=0) = 0,$$

where $\eta(x)$ is a randomly fluctuating function in $0 < x < L$ defined by $\eta(x) = n^2(x) - 1$.

Typically in the studies of quantum mechanical or optical systems [1,6], the potential or refractive index within a random medium is assumed to fluctuate around the potential or index level of its surrounding medium. When the surrounding medium is a vacuum, this means $\eta(x)$ has zero mean, i.e., $<\eta(x)> = 0$. However, in this paper, we consider an optical random medium whose mean refractive index, $n_0$ ($=<n(x)>$), does not match with the index of its surrounding medium. (See Fig. 1.) Therefore, the effect of this refractive index mismatch at $x = 0$ and $L$ needs to be taken into account in the analysis of the reflection statistics.

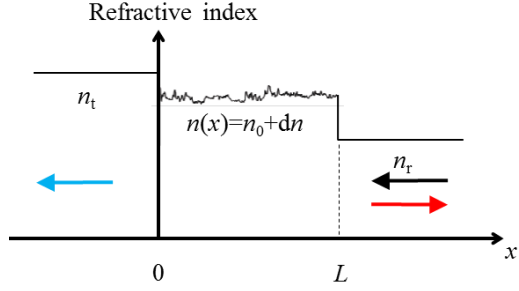

**Fig. 1.** 1D Light scattering by a random optical medium is described. Light is incident on the medium from the right (black arrow), reflected (red arrow) and transmitted (blue arrow). Due to index mismatch at *x*=0 and *L*, the light experiences an interference caused by boundary reflections. $n_r$ and $n_t$ are the refractive indices of the surrounding

To include the effects of an index mismatch, we view the random medium of Fig. 1 as an optical medium with a constant index that is perturbed by weak disorder. Then, the sample medium index can be written as $n(x) = n_0 + \Delta n(x)$, where $n_0 = <n(x)>$ and $\Delta n$ is a deviation around $n_0$. Accordingly, we can rewrite the random function $\eta(x)$ in Eq. (2a) as a sum of its mean and fluctuation:

$$\frac{dR(L)}{dL} = 2ikR(L) + \frac{ik}{2}[(n_0^2 - 1) + n_0\eta_d][1 + R(L)]^2, \qquad (2b)$$

where $\eta_d = 2\Delta n + O(\Delta n^2)$. From here, $O(\Delta n^2)$ in $\eta_d$ is ignored since the weak disorder limit ($\Delta n \ll n_0$) is applied. Now, we consider $R(L)$ in Eq. (2b) as a sum of two contributions: reflection due to the boundary index mismatch ($n_0 \neq n_{out}$) and a perturbation due to weak disorder $\Delta n$. Accordingly, we rewrite $R(L)$: $R(L) = R_s(L) + \Delta R(L)$, where $R_s(L)$ is a deterministic reflection coefficient due to the interference between two boundaries ($x = 0$ and $L$) and $\Delta R(L)$ is the remaining contribution due to $\Delta n$. The differential equation of $R_s(L)$ is given in Eq. (2c) by $\Delta n \to 0$ in Eq. (2b). Then, the stochastic equation for $\Delta R(L)$ can be derived as in Eq. (2d) by subtracting Eq. (2c) from Eq. (2b) [6]:

$$\frac{dR_s(L)}{dL} = 2ikR_s + \frac{ik}{2}(n_0^2 - 1)[1 + R_s]^2, \qquad (2c)$$

$$\frac{d\Delta R(L)}{dL} = 2ik\Delta R + \frac{ik}{2}n_0\eta_d[1 + R_s + \Delta R]^2 \\ + \frac{ik}{2}(n_0^2 - 1)[2\Delta R(1 + R_s) + \Delta R^2], \qquad (2d)$$

where $R_s(L = 0) = 0$ and $\Delta R(L = 0) = 0$. Eq. (2c) is a simple interference problem with a thin slab geometry of two boundaries ($x = 0$ and $L$). $R_s$ is well described both as a perturbative form and as a closed form [11]. In section 2.3, the perturbative form is used to calculate the mean and STD of the reflectance.

*2.2 Weak disorder with short range spatial correlation: different averages*

Before obtaining the mean and STD of reflectance, it is necessary to consider the stochastic behavior of a sample refractive index. This gives us the ability to calculate the moments of the complex reflection amplitude difference $\Delta R$. To describe the stochastic properties of the refractive index, we introduce a spatially correlated disorder into the optical medium in Fig. 1. Specifically, we adopt an Ornstein-Uhlenbeck stochastic process and establish statistical properties of $\eta_d$ (= $2\Delta n$) in Eq. (2b) [6]:

$$<\eta_d> = 0,$$
$$<\eta_d(x)\eta_d(x')> = \Delta^2 \exp(-|x-x'|/l_c), \tag{3a}$$

where $l_c$ is a correlation length of $\Delta n$ and $\Delta^2 = 4<\Delta n^2>$. In this paper, we assume a short range correlation, $kl_c < 1$. When $\eta_d$ satisfies Eq. (3a), its differentiation formula is derived for a functional of $\eta_d$ [6]:

$$\frac{\partial}{\partial x}<\eta_d(x)\Phi[\eta_d(x)]> = \left\langle \eta_d(x)\frac{\partial}{\partial x}\Phi[\eta_d(x)]\right\rangle - \frac{1}{l_c}<\eta_d(x)\Phi[\eta_d(x)]>, \tag{3b}$$

where $\Phi$ is a functional of $\eta_d$.

Now differential equations for any moments of $\Delta R(L)$ can be derived based on Eq. (3b). Therefore, the mean and STD of reflectance can also be calculated (see the next section). To assist the calculation, we first calculate $<\eta_d\Delta R>$, which is a stochastic contribution of the mean and STD in the leading order of $\Delta^2$. Using Eqs. (2d) and (3b) with $\Phi = \Delta R$ and keeping $\Delta R$ up to $O(\eta_d^1)$, a formal expression of $<\eta_d\Delta R>$ up to $O(\Delta^2)$ can be written (see the appendix):

$$<\eta_d\Delta R> = \frac{i}{2}n_0 k l_c \Delta^2 \left(\frac{1}{1-2ikl_c\beta}\right)[1-l_c\frac{\partial}{\partial L}](1+R_s)^2, \tag{4}$$

where $\beta = 1 + (n_0^2 - 1)(1 + R_s)/2$. Here, we introduce a phase transformation which simplifies Eq. (2d) by removing $O(\eta_d^0)$ and makes the perturbative calculation straightforward in the next section:

$$\Delta R(L) = \Delta Q(L) \cdot e^{2ik\alpha(L)}, \tag{5}$$

where $\alpha = \int_0^L \beta(L')dL'$. The same relationship exists between $R$ and $Q$, and $R_s$ and $Q_s$. The phase term, $e^{2ik\alpha}$, is used to separate the rapid phase oscillation caused by $2ikR$ in Eqs. (2a-d). When the constant part of the random potential $\eta$, $(n_0^2 - 1)$, vanishes, the phase factor $e^{2ik\alpha}$ in Eq. (5) becomes $e^{2ikL}$, which is a mere phase delay due to a round trip within $[0, L]$ in a vacuum. In this $Q$ representation, Eq. (2d) becomes:

$$\frac{d\Delta Q}{dL} = \frac{i}{2}n_0 k \eta_d e^{-2ik\alpha}[1 + R_s + \Delta Q e^{2ik\alpha}]^2 + \frac{i}{2}k(n_0^2 - 1)(\Delta Q)^2 e^{2ik\alpha}$$
$$= \frac{i}{2}n_0 k \eta_d e^{-2ik\alpha}[1 + R_s]^2 + O(\eta_d^2). \tag{6}$$

The leading order of $\eta_d$ in Eq. (6) is $\eta_d^1$, whereas that of Eq. (2d) is $\eta_d^0$. By using Eq. (5), the expression corresponding to Eq. (4) can be written again up to $O(\Delta^2)$:

$$<\eta_d\Delta Q> = \frac{i}{2}n_0 k l_c \Delta^2 \left(\frac{e^{-2ik\alpha}}{1-2ikl_c\beta}\right)[1-l_c\frac{\partial}{\partial L}](1+R_s)^2. \tag{7}$$

Now, the mean and STD of reflectance can be readily derived based on Eq. (7) in the $Q$ representation.

*2.3 Mean and STD of the reflectance calculations*

In this section, we calculate the mean and STD of reflectance perturbatively up to $O(\Delta^2)$. For a perturbative expansion, the mean and variance can be decomposed into deterministic terms and disorder terms by using $R(L) = R_s(L) + \Delta R(L)$.

$$\begin{aligned}<r> &= <(R_s + \Delta R)(R_s + \Delta R)^*> \\ &= r_s + R_s^* <\Delta R> + c.c. + <|\Delta R|^2>,\end{aligned} \quad (8a)$$

$$\begin{aligned}\sigma^2(r) &= <r^2> - (<r>)^2 \\ &= 2r_s[<|\Delta R|^2> - <\Delta R^*><\Delta R>] \\ &\quad + R_s^2[<\Delta R^{*2}> - <\Delta R^*>^2] + c.c. \\ &\quad + 2R_s[<\Delta R^*|\Delta R|^2> - <\Delta R^*><|\Delta R|^2>] + c.c. \\ &\quad + <|\Delta R|^4> - <|\Delta R|^2>^2,\end{aligned} \quad (8b)$$

where $r = |R|^2 = |Q|^2$ and $r_s = |R_s|^2 = |Q_s|^2$. The moments of $\Delta R$, $f_{m,n} = <\Delta R^m \Delta R^{*n}>$, satisfy $f_{m,n} \sim \delta_{m,n}$ in the weak disorder limit as discussed elsewhere [6]. Then, Eqs. (8a) and (8b) can be simplified in the $R$ and $Q$ representations:

$$<r> \approx r_s + <|\Delta R|^2> = r_s + <|\Delta Q|^2>, \quad (9a)$$

$$\begin{aligned}\sigma^2(r) &\approx 2r_s <|\Delta R|^2> + <|\Delta R|^4> - <|\Delta R|^2>^2 \\ &= 2r_s <|\Delta Q|^2> + <|\Delta Q|^4> - <|\Delta Q|^2>^2.\end{aligned} \quad (9b)$$

From here, we use the $Q$ representation that establishes simpler perturbative expansions. Among three terms in Eq. (9b), $r_s<|\Delta Q|^2>$ is $O(\Delta^2)$ and the rest of the terms have higher orders in $\Delta^2$. Therefore, $\sigma^2(r) = 2r_s<|\Delta Q|^2>$ in $O(\Delta^2)$. Now Eqs. (9a) and (9b) can be calculated by integrating the differential form of $<|\Delta Q|^2>$:

$$\frac{d<|\Delta Q|^2>}{dL} = <\left(\frac{d\Delta Q}{dL}\right)\cdot \Delta Q^*> + <\Delta Q \cdot \left(\frac{d\Delta Q^*}{dL}\right)>. \quad (10)$$

Using Eq. (6) and keeping the order of disorder only up to $\Delta^2$, $<|\Delta Q|^2>$ is expressed:

$$\begin{aligned}<|\Delta Q|^2> &= -\frac{i}{2}n_0 k \int_0^L dL'[e^{2ik\alpha^*}(1+R_s^*)^2 <\eta_d \Delta Q>] + c.c. \\ &= \frac{1}{4}(n_0 k)^2 l_c \Delta^2 \int_0^L dL' e^{2ik(\alpha^*-\alpha)}\left(\frac{1}{1-2ikl_c\beta}\right)(1+R_s^*)^2[1-l_c\frac{\partial}{\partial L}](1+R_s)^2 + c.c.\end{aligned} \quad (11)$$

Here, it is important to note that $\sigma^2(r) = 2r_s<|\Delta Q|^2>$ in Eq. (9b) is valid only when $R_s$ is significantly greater than $\Delta R$. For example, $\sigma^2(r) = 2r_s<|\Delta Q|^2>$ is invalid under a complete destructive interference condition which reduces $R_s$ to zero for some periodic $k$'s with a given $L$. At these periodic $k$'s, $\sigma^2(r)$ tends to the expression in the index matched system [6]: $\sigma^2(r) = <|\Delta Q|^4> - <|\Delta Q|^2>^2 \sim <\Delta n^2>^4 l_c^2$.

To perform the integration in Eq. (11), we consider $R_s$, which is simply a solution to Eq. (2c). A closed form of $R_s$ is not suitable for performing the integration due to its fractional form. (See 4.27 and 4.30 in [11]) Therefore, we expand $R_s$ as a sum of all successive rays that are reflected and transmitted at $x = 0$ and $L$ and approximate $R_s$ by keeping the two lowest order terms [11]:

$$R_s \approx R_s^{(2)} = A_0 + A_1 e^{2in_0 kL}, \tag{12}$$

where $A_0 = (n_r - n_0)/(n_r + n_0)$ and $A_1 = 4 n_r n_0/(n_r + n_0)^2 \cdot (n_0 - n_t)/(n_0 + n_t)$ ($|A_0|$ and $|A_1| < 1$). We also define $B_0$ as $B_0 = A_0 + 1$ for the calculations below. The approximation in Eq. (12) is accurate when reflections at the boundaries are not very strong ($|A_0|$ and $|A_1| \ll 1$) as in biological samples. In the appendix, the integration of Eq. (11) is performed expanding the integrand in terms of $A_1$ with a periodic condition of $k$: $2 n_0 k L = \pi N$ ($N$ = even integers). Then, $<|\Delta Q|^2>$ is approximated in $O(\Delta^2)$:

$$<|\Delta Q|^2> = \frac{1}{2}(n_0 k_N)^2 l_c \Delta^2 \frac{L}{1+4(\rho k_N l_c)^2} F(n_0, n_r, n_t), \tag{13}$$

where $2 n_0 k_N L = \pi N$ ($N = 2, 4, 6\ldots$), $\rho = 1 + (n_0^2 - 1) B_0 / 2$, $F(n_0, n_r, n_t) = e^{\frac{(n_0^2-1)A_1}{n_0}} I_0(\left|\frac{(n_0^2-1)A_1}{n_0}\right|) \times B_0^4$, and $I_0$ is a Modified Bessel function of the first kind. Accordingly, the mean and STD in Eqs. (9a) and (9b) are expressed by Eq. (13):

$$<r> \approx r_s + <|\Delta Q|^2>$$
$$= (A_0^2 + A_1^2 + 2 A_0 A_1 \cos x_N) + \frac{1}{2}(n_0 k_N)^2 l_c \Delta^2 \frac{L}{1+4(\rho k_N l_c)^2} F(n_0, n_r, n_t) \tag{14}$$
$$= (A_0 + A_1)^2 + \frac{1}{2}(n_0 k_N)^2 l_c \Delta^2 \frac{L}{1+4(\rho k_N l_c)^2} F(n_0, n_r, n_t),$$

$$\sigma^2(r) \approx 2 r_s <|\Delta Q|^2>$$
$$= 2(A_0^2 + A_1^2 + 2 A_0 A_1 \cos x_N) \times \frac{1}{2}(n_0 k_N)^2 l_c \Delta^2 \frac{L}{1+4(\rho k_N l_c)^2} F(n_0, n_r, n_t) \tag{15}$$
$$= (A_0 + A_1)^2 \cdot (n_0 k_N)^2 l_c \Delta^2 \frac{L}{1+4(\rho k_N l_c)^2} F(n_0, n_r, n_t),$$

where $x_N = 2 n_0 k_N L = \pi N$ ($N = 2, 4, 6\ldots$).

Finally, we compare the above theoretical result with direct numerical simulation data obtained from iteration Eqs. (2a) and (2b) with the initial condition R(L=0)=0. For the simulation, 10,000 realizations of correlated random noise, $\eta$, were generated for each parameter set ($\Delta$, $l_c$, $n_0$, $k$ and $L$). Then, the corresponding reflectance is calculated with those realizations based on Eq. (2a). During the calculation, the Riccati form of Eq. (2a) was linearized for stable numerical integration. In Fig. 2 and Fig. 3, STDs based on Eq. (15) are plotted with the exact numerical simulation data in two disorder parameter spaces,

$<\Delta n^2>^{1/2}$ and $l_c$, to verify the accuracy of the analytical calculations. In Fig. 2, we observe a clear difference between index mismatched systems and a matched system in terms of the order of $<\Delta n^2>^{1/2}$ in STD (see section 3.).

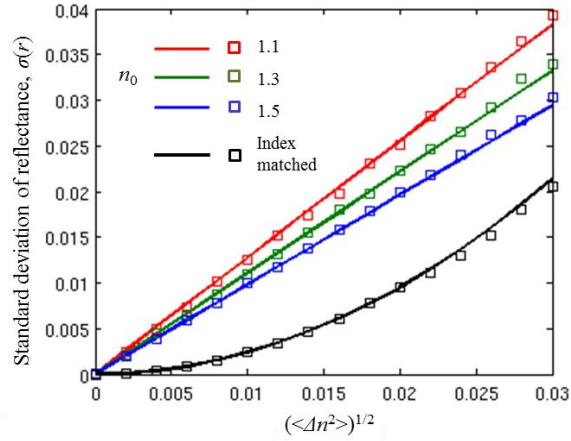

**Fig. 2.** STD vs. $<\Delta n^2>^{1/2}$ (=$\Delta/2$) of reflection statistics. A set of parameters is: $n_r$=1, $n_t$=1.5, $n_0$=1.1~1.5, $l_c$=20 nm, $\lambda$=550 nm, and $2L$=$(1/2\cdot\lambda/n_0)\cdot 20$. The colored solid lines are theoretical values based Eq. 15 and the colored markers are simulational data points. The black markers and black solid line are simulational data points and theoretical calculation based on Eq.16b with $n_r$=1, $n_t$=1.0 and $n_0$=1.0 (an index matched system) shown for comparison.

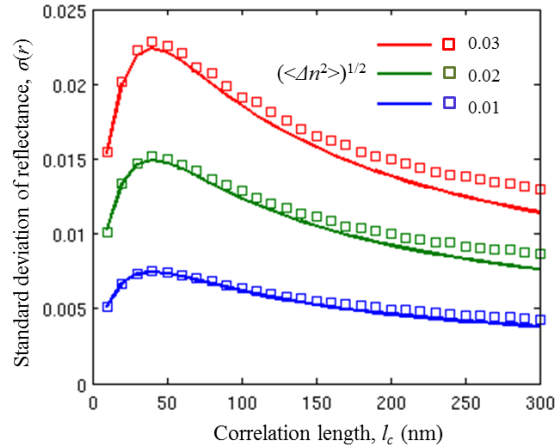

**Fig. 3.** STD vs. $l_c$ of reflection statistics. A set of parameters is: $n_r$=1, $n_t$=1.5, $n_0$=1.3, $<\Delta n^2>^{1/2}$=0.01-0.03, $\lambda$=650 nm, and $2L$=$(1/2\cdot\lambda/n_0)\cdot 16$. The solid lines are theoretical values based Eq. 15 and the markers are simulational data points.

## 3. Discussion: analysis of mean and STD of reflectance and applications

Reflection statistics have been well established for index matched systems. The mean and STD of reflectance for index matched systems exhibit the same behavior in the weak disorder limit (4.56 in [6]):

$$<r_{mat}> = L/\xi$$
$$= \frac{1}{2}k^2 l_c \Delta^2 \frac{L}{1+4(kl_c)^2}, \tag{16a}$$

$$\sigma(r_{mat}) \approx <r_{mat}>, \tag{16b}$$

where the inverse of localization length is : $\xi^{-1} = 2^{-1}k^2 l_c \Delta^2/[1+(2kl_c)^2]$ and $L \ll \xi$.

$<r>$ of a mismatched system in Eq. (14) has essentially the same stochastic behavior as Eq. (16a). This is because the interference between boundary reflection and disorder scattering, $Q_s^* \Delta Q + c.c.$, in Eq. (8a) vanishes upon ensemble averaging. Therefore, Eq. (14) is an additive combination of deterministic boundary interference $r_s$ and stochastic contribution $<|\Delta Q|^2>$. Note that $<|\Delta Q|^2>$ exhibits the same behavior as in (16a):

$$<|\Delta Q|^2> \sim L/\xi \sim \frac{1}{2}(n_0 k_N)^2 l_c \Delta^2 \frac{L}{1+4(\rho k_N l_c)^2}.$$

Here, the correction to the wave number ($k \to n_0 k$) is due to the mean refractive index, and the rest of the $n_0$ dependent terms, $\rho$ and $F(n_0)$, in Eq. (14) reduce to one as $n_0 \to 1$.

*On the other hand, $\sigma(r)$ in Eq. (15) is described as a multiplicative combination of $r_s$ and $<|\Delta Q|^2>$. This leads to a drastic deviation of $\sigma(r)$ from $\sigma(r_{mat})$ in terms of the order of disorder parameters as in Eqs. (17a) and (17b). The orders of $\Delta$ and $l_c$ decrease twice from $\sigma(r_{mat})$ to $\sigma(r)$, and $\sigma(r)$ has a deterministic term $(2r_s)^{1/2}$ as a prefactor that modulates the amplitude of $\sigma(r)$ as demonstrated in Fig. 2. The origin of this deviation is the interference between boundary reflection and disorder scattering, $Q_s^* \Delta Q + c.c.$ in $r$ of Eq. (8b).*

$$\sigma(r_{mat}) \sim \Delta^2 \frac{k^2 l_c L}{1+4(kl_c)^2} \sim \Delta^2 (kl_c)^1 \cdot (kL)^1, \tag{17a}$$

$$\sigma(r) \approx \sqrt{2r_s}\sqrt{<|\Delta Q|^2>}$$
$$\sim \sqrt{2(A_0^2 + A_1^2 + 2A_0 A_1 \cos x_N)} \cdot \Delta \sqrt{\frac{k_N^2 l_c L}{1+4(n_0 k_N l_c)^2}} \tag{17b}$$
$$\sim \sqrt{2r_s} \cdot \Delta^1 (k_N l_c)^{1/2} (k_N L)^{1/2},$$

where $x_N = 2n_0 k_N L = \pi N$ ($N = 2, 4, 6...$).

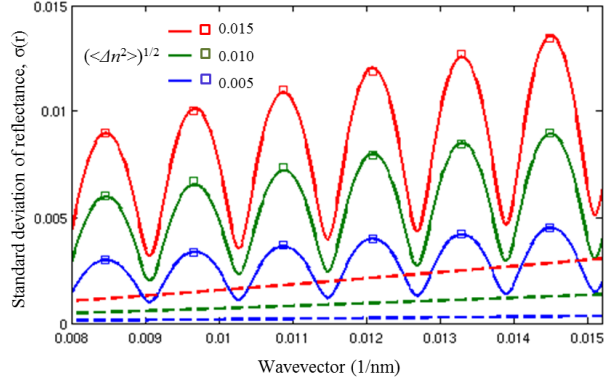

**Fig.4.** $\sigma(r)$ vs. $k$ of the reflection statistics. A set of parameters for refactive indices for outside meduun and the sample: $n_r$ =1, $n_t$ =1.5, $n_0$=1.3, $l_c$=20 nm, $<\Delta n^2>^{1/2}$=0.005-0.015, $\lambda$=400-800 nm, and $L=2\mu$. The solid lines are simulational data and the markers are theoretical calculation based Eq. 15. For the theoretical calculation, a condition $2L=(1/2\cdot\lambda_N/n_0)\cdot N$ ($N$= even integers) was used, where $k_N=2\pi/\lambda_N$. The dashed lines are theoretical values based on Eq.16b with $n_r$ =1, $n_t$ =1.0, $n_0$ =1.0 (an index matched system) and the same

*Enhancements of the STD of the reflectance ($\sigma(r)$)* : To analyze the effect of $(2r_s)^{1/2}$ and decreased orders of $\Delta$ and $l_c$ in Eq. (17b), STDs of several $\Delta$'s are presented for both mismatched and matched systems in Fig. 4. First, $\sigma(r)$'s show a pseudo-periodic and 'amplified' behavior in the $k$ domain (solid lines of simulation data), whereas $\sigma(r_{mat})$'s show a small monotonic increase as $k$ increases. The periodic modulation is driven by the boundary interference $(2r_s)^{1/2}$ and, therefore, $\sigma(r)$ is maximally amplified around maxima of $(2r_s)^{1/2}$. Second, the decreased order of disorder parameters ($\Delta^2(kl_c) \rightarrow \Delta(kl_c)^{1/2}$) improve sensitivity of $\sigma(r)$ to disorder parameters ($\Delta$ and $l_c$) in the weak disorder and short range correlation limits. In Fig. 4, we find that $\sigma(r)/\sigma(r_{mat}) > 10$ for $<\Delta n^2>^{1/2}$=0.005 and $(\sigma(r)_{0.010}-\sigma(r)_{0.005})/(\sigma(r_{mat})_{0.010}-\sigma(r_{mat})_{0.005}) \sim 5$ around maxima of $(2r_s)^{1/2}$. This clearly is a an example of the enhancement of the reflectance by the value of the refractive index mismatch.

*Applications:* This periodic amplification and enhanced sensitivity of $\sigma(r)$ has a strong potential application in the context of disorder information extraction efficiently from weakly spatial disorder fluctuations imbedded in a large refractive index media, such as biological cells. In particular, it can be a physical basis for various techniques in biophotonics since the index mismatch condition and weak disorder limit are readily applicable to biological cells, which have mean refractive index $n_0 \sim 1.3 – 1.5$ and spatial refractive index fluctuations $<\Delta n^2>^{1/2} \sim 0.01 – 0.1$. The light backscattering by weakly disordered systems such as biological cells can be approximated as a quasi-1D parallel multichannel problem [12]. It was recently demonstrated that quasi-1D multichannel backscattering statistics of biological cells are sensitive to changes in nanoscale disorder properties [13-16]. Based on the first order Born approximation, the reflection statistics were also theoretically analyzed in 3D backscattering, which can be applied to 1D case as well. [17]. It is demonstrated that the applications of the light backscattering statistics can be applied for early pre-cancer screening by detecting changes in the refractive index fluctuation in cells, corresponding to the progress of carcinogenesis in different types of cancer. By enhancing the STDs of reflected signals via controlling $(2r_s)^{1/2}$ factor, we expect to achieve a high detection sensitivity to a waek disorder parameters of the nanoscale refractive index fluctuations in biological cells.

## 4. Conclusions

In conclusion, we performed a theoretical analysis to understand the effects of a refractive index mismatch between a weakly disordered optical sample and its surrounding medium on reflection statistics. By separating the interference induced by sample boundaries and the disorder scattering, we show that the mean of the reflectance $<r>$ in a mismatched system has the same stochastic behavior as that of a matched system. On the other hand, the STD of a mismatched system $\sigma(r)$ has lower orders of disorder parameters compared to that of a matched system, and the boundary interference term $(2r_s)^{1/2}$ functions as a prefactor. In particular, the prefactor $(2r_s)^{1/2}$ causes a periodic amplification of the disorder signal in the spectral domain. The origin of this difference between $\sigma(r)$ and $\sigma(r_{mat})$ is the interference between the boundary reflection and disorder scattering in $\sigma^2(r)$. From a technological point of view, the boundary index mismatch condition provides a handle for enhancing the signal from the disorder part of the refractive index fluctuations buried in a large uniform refractive index media. The intracellular mass density fluctuations or refractive index fluctuations increase with progress of carcinogenesis, therefore, probing weak structural disorder parameters in biological cells with high detection sensitivity by tuning the index mismatch condition has strong potential applications in early cancer screening, as well as other biophotonics applications that probe mass density variations or refractive index fluctuations.

## 5. Appendix: Derivations of Eqs. (4) and (13)

*5.1 Derivation of Eq. (4)*

First, to derive Eq. (4), we use Eq. (3b) under the assumption of an Ornestein-Uhlenbeck stochastic process. Then, a differentiation formula for $\Delta R$ can be established:

$$[1+l_c\frac{\partial}{\partial L}]<\eta_d\Delta R>=\left\langle\eta_d l_c\left(\frac{\partial}{\partial L}\Delta R\right)\right\rangle. \quad (A.1)$$

Using Eq. (2d) and maintaining the order of $\eta_d$ only up to $O(\eta_d^2)$ on the right side of Eq. (A.1), we obtain:

$$[1+l_c\frac{\partial}{\partial L}]<\eta_d\Delta R>$$
$$=\left\langle\eta_d l_c\left(2ik\Delta R+\frac{ik}{2}n_0\eta_d[1+R_s+\Delta R]^2+\frac{ik}{2}(n_0^2-1)[2\Delta R(1+R_s)+\Delta R^2]\right)\right\rangle \quad (A.2)$$
$$=2ikl_c\beta<\eta_d\Delta R>+\frac{i}{2}n_0kl_c\Delta^2(1+R_s)^2,$$

where $\beta = 1 + (n_0^2 - 1)(1 + R_s)/2$. Multiplying Eq. (A.2) by $[1-l_c\frac{\partial}{\partial L}]$, we obtain:

$$<\eta_d \Delta R> - l_c^2 \frac{\partial^2}{\partial L^2} <\eta_d \Delta R>$$
$$= [1 - l_c \frac{\partial}{\partial L}] \left( 2ikl_c \beta <\eta_d \Delta R> + \frac{i}{2} n_0 k l_c \Delta^2 (1+R_s)^2 \right) \quad (A.3)$$
$$= 2ikl_c \beta <\eta_d \Delta R> - 2ikl_c^2 \frac{\partial}{\partial L} \beta <\eta_d \Delta R> + \frac{i}{2} n_0 k l_c \Delta^2 [1 - l_c \frac{\partial}{\partial L}](1+R_s)^2.$$

By rewriting Eq. (A.3) in orders of $kl_c$, Eq. (A.3) is expressed:

$$(1 - 2ikl_c \beta) <\eta_d \Delta R> + \hat{O} <\eta_d \Delta R>$$
$$= \frac{i}{2} n_0 k l_c \Delta^2 [1 - l_c \frac{\partial}{\partial L}](1+R_s)^2, \quad (A.4)$$

where $\hat{O}$ is an operator whose lowest order is $(kl_c)^2$ and the lowest order of $kl_c$ in $<\eta_d \Delta R>$ is $(kl_c)^1$. Approximating Eq. (A.4) by ignoring $\hat{O}$ in the short spatial correlation limit, we obtain:

$$<\eta_d \Delta R> = \frac{i}{2} n_0 k (\Delta^2 l_c) \left( \frac{1}{1 - 2ikl_c \beta} \right) [1 - l_c \frac{\partial}{\partial L}](1+R_s)^2. \quad (A.5)$$

*5.2 Derivation of Eq. (13)*

Secondly, to derive $<|\Delta Q|^2>$ in Eq. (13), we first insert Eq. (12) into Eq. (11) and change the variable $L$ by $x = 2n_0 k L$. This gives:

$$<|\Delta Q|^2> = -\frac{i}{2} n_0 k \int_0^L dL' [e^{2ik\alpha^*} (1+R_s^*)^2 <\eta_d \Delta Q>] + c.c.$$
$$= \frac{1}{8} (n_0 k l_c) \Delta^2 \int_0^x dx' e^{2ik(\alpha^* - \alpha)} \left( \frac{1}{1 - 2ikl_c \rho - ikl_c (n_0^2 - 1) A_1 e^{ix'}} \right) \quad (A.6)$$
$$\times (B_0 + A_1 e^{-ix'})^2 [1 - 2n_0 k l_c \frac{\partial}{\partial L}] (B_0 + A_1 e^{ix'})^2 + c.c.,$$

where $\rho = 1 + (n_0^2 - 1)B_0/2$. The exponent of $e^{2ik(\alpha^* - \alpha)}$ in Eq. (A.6) can be integrated:

$$2ik(\alpha^* - \alpha) = -ik(n_0^2 - 1) \int_0^L dL' (R_s - R_s^*)$$
$$= -ik(n_0^2 - 1) \int_0^L dL' [(A_0 + A_1 e^{2in_0 kL}) - (A_0 + A_1 e^{-2in_0 kL})] \quad (A.7)$$
$$= \frac{(n_0^2 - 1)A_1}{n_0} (1 - \cos x).$$

Eq. (A.6) can, then, be expanded in terms of $A_1$ to perform the integral because $A_1 < 1$. By maintaining the order of $A_1$ only up to $(A_1)^1$, we obtain:

$$<|\Delta Q|^2>$$

$$= \frac{1}{8} n_0 k l_c \Delta^2 e^{\frac{(n_0^2-1)A_1}{n_0} x} \int_0^{x} dx' e^{\frac{-(n_0^2-1)A_1}{n_0} \cos x'} \left( 1 + \frac{i(kl_c)(n_0^2-1)A_1 e^{ix'}}{1-2i\rho k l_c} \right)$$

$$\times (B_0 + A_1 e^{-ix'})^2 [1 - l_c \frac{\partial}{\partial L}](B_0 + A_1 e^{ix'})^2 + c.c.$$

$$= \frac{1}{8} n_0 k l_c \Delta^2 e^{\frac{(n_0^2-1)A_1}{n_0} x} \int_0^{x} dx' e^{\frac{-(n_0^2-1)A_1}{n_0} \cos x'} \frac{1}{1+4(\rho l_c)^2}$$

$$\times \begin{bmatrix} 2B_0^4 \\ +(B_0^3 A_1)\left( 8(1 + 2n_0 \rho(kl_c)^2) - \frac{8B_0(n_0^2-1)\rho(kl_c)^2}{(1+4(\rho kl_c)^2)} \right) \cos x' \\ +(B_0^3 A_1)\left( 8n_0(kl_c) - \frac{2B_0(n_0^2-1)(kl_c)(1-4(\rho kl_c)^2)}{(1+4(\rho kl_c)^2)} \right) \sin x' \end{bmatrix} \quad (A.8)$$

The integral in Eq. (A.8) can be performed by a set of known integral formulas in Eq. (A.9) for $x = 2n_0 k_N L = N\pi$ when $N$ is an even integer:

$$\int_0^{\pi} dx \cdot e^{p \cos x} \cos mx = \pi \cdot I_m(|p|),$$

$$\int_0^{N\pi} dx \cdot e^{p \cos x} \cos mx = N\pi \cdot I_m(|p|), \quad (A.9)$$

$$\int_0^{2\pi} dx \cdot e^{p \cos x} \sin mx = 0,$$

where $I_m$ is a modified Bessel function of the first kind and $m$ is an integer. Then, $<|\Delta Q|^2>$ can be expressed up to $O(A_1^1)$ and $O(\Delta^2)$:

$$<|\Delta Q|^2> = \frac{1}{2} n_0^2 k_N^2 l_c \Delta^2 \frac{L}{1+4(\rho k_N l_c)^2} e^{\frac{(n_0^2-1)A_1}{n_0}}$$

$$\times \begin{bmatrix} B_0^4 I_0\left( \left| \frac{(n_0^2-1)A_1}{n_0} \right| \right) \\ + 4B_0^3 A_1 \left\{ 1 + 2n_0 \rho(k_N l_c)^2 - \frac{(n_0^2-1)B_0 \rho(k_N l_c)^2}{1+4(\rho k_N l_c)^2} \right\} I_1\left( \left| \frac{(n_0^2-1)A_1}{n_0} \right| \right) \end{bmatrix}. \quad (A.10)$$

By ignoring the second term (leading order, $A_1^1$) in the bracket, Eq. (A.10) can be approximated:

$$<|\Delta Q|^2> = \frac{1}{2}(n_0 k_N)^2 l_c \Delta^2 \frac{L}{1+4(\rho k_N l_c)^2} F(n_0, n_r, n_t), \quad (A.11)$$

where $F(n_0, n_r, n_t) = e^{\frac{(n_0^2-1)A_1}{n_0}} I_0(\left|\frac{(n_0^2-1)A_1}{n_0}\right|) B_0^4$.

**Acknowledgements**

This work was supported by NIH grants R01 EB003682, R01 CA155284, and R01 CA165309.